\documentclass[a4paper,11pt]{article}

\usepackage[latin1]{inputenc}
\usepackage[english]{babel}
\usepackage{makeidx} 
\usepackage{amsfonts,amssymb,amsmath,amsthm}
\usepackage{pdfsync}
\usepackage{color}
\usepackage{nicefrac}
\usepackage{hyperref}
\hypersetup{colorlinks=true,citecolor=black,filecolor=black,linkcolor=blue,urlcolor=black,pdftex}

\newtheorem{prop}{Proposition}[section]
\newtheorem{theo}[prop]{Theorem}
\newtheorem{lemm}[prop]{Lemma}

\makeatletter
\@addtoreset{equation}{section}

\makeatother

\renewcommand{\u}{\tilde{u}}
\newcommand{\eps}{\varepsilon}
\newcommand{\R}{\mathbb{R}}
\newcommand{\C}{\mathbb{C}}
\newcommand{\N}{ \mathbb{N}}
\newcommand{\E}{ \mathcal{E}}
\newcommand{\F}{ \mathcal{F}}
\newcommand{\U}{ \mathcal{U}}
\newcommand{\V}{ \mathcal{V}}
\newcommand{\Y}{ \tilde{Y}}
\renewcommand{\O}{ \mathcal{O}}

\renewcommand{\L}{\Lambda}

\renewcommand{\u}{\tilde{u}}
\renewcommand{\v}{\tilde{v}}
\newcommand{\w}{\tilde{w}}

\newcommand{\be}{\begin{equation}}
\newcommand{\ee}{\end{equation}}
\newcommand{\ben}{\begin{equation*}}
\newcommand{\een}{\end{equation*}}
\newcommand{\ba}{\begin{eqnarray}}
\newcommand{\ea}{\end{eqnarray}}
\newcommand{\ban}{\begin{eqnarray*}}
\newcommand{\ean}{\end{eqnarray*}}

\begin{document}

\title{Segregation  and symmetry breaking of strongly coupled two-component Bose-Einstein condensates in a harmonic trap}
         
\author{ {\Large Jimena Royo-Letelier} \\\\  { \small Laboratoire de Math\'ematiques de Versailles} \\ { \small Universit\'e Versailles-Saint-Quentin-en-Yvelines} \\ { \small 45
avenue des \'Etats-Unis, 78035 Versailles C\'edex, France}  \\ { \small jimena.royo-letelier@uvsq.fr}  }
\date{\today}

\maketitle

\begin{abstract}
{We study ground states of two-component condensates in a harmonic trap. We prove that in the strongly coupled and weakly interacting regime, the two components segregate while a symmetry breaking occurs. More precisely, we show that when the intercomponent coupling strength is very large and both intracomponent coupling strengths are small, each component is close to the positive or the negative part of a second eigenfunction of the harmonic oscillator in $\R^2$. As a result, the supports of the components approach complementary half-spaces, and they are not radially symmetric. }
\end{abstract}


\section{Introduction}
\label{intro}

	A two-component Bose-Einstein condensate (BEC) is described in terms of two wave functions $u$ and $v$, respectively representing the first and the second component. The energy of a trapped two-dimensional two-component BEC is given for $g=(g_1,g_2,g_{12})$  by  

\ben
	\E_g(u, v) =  \frac12 \int_{\R^2}  \Big\{  |\nabla u|^2 + V |u|^2 + \frac12 g_1 |u|^4 \Big\} + \Big\{ |\nabla v|^2 + V |v|^2 + \frac12 g_2 |v|^4 \Big\} +   g_{12} | u|^2 |v|^2  \,.
 \een
 
Here $g_{12}$ is the intercomponent coupling strength, and $g_1$ (respectively $g_2$) is the intracomponent coupling strength of the first (respectively second) component. We are interested in the properties of the ground state of $\E_g$ when $g_{12}$ goes to infinity. We assume that both components have repulsive internal interactions, so 
 
 \be \label{sharon}
 	g_1\geq 0   \qquad \text{ and } \qquad g_2\geq 0 \,,
 \ee
 
 and that the trapping potential is the harmonic function
 
 \be \label{jones}
	V(x) = \omega^2 |x|^2 \,, \quad \omega >0 \,.
 \ee
  
We consider then $(u_g,v_g)$ as a  minimizer of $\E_g$ over the class of finite energy pairs with constrained mass
 
 \ben
	X = \Big\{ (u,v) \,;\, u,v \in H^1_V(\R^2;\C)  \,,\,  \| u\|_{L^2(\R^2)} =1 ,   \| v\|_{L^2(\R^2)} =1 \Big\} \,,
 \een
 
 where
 
 \ben
 	H^1_V(\R^2;\C) = \Big\{ u \in H^1(\R^2;\C)  \,;\,  \int_{\R^2} V |u|^2 < \infty \Big\} \,. 
 \een
 
 The pair $(u_g,v_g)$ satisfies the system of coupled Gross-Pitaevskii equations
 
 \be \label{eq:sc}
	 \left. \bigg \{
	\begin{array}{lll}
		-\Delta u + V u  + g_1 |u|^2u+ g_{12} |v|^2 u &=& \lambda u \\
		-\Delta v + V v  + g_2 |v|^2v + g_{12}  |u|^2 v &=& \mu v \,
	\end{array}  \right.
\ee
 
in $\R^2$, where $\lambda$ and $\mu$ are the Lagrange multipliers associated with the mass constraints.  \\

The system (\ref{eq:sc}) is a particular case of nonlinear elliptic systems with competition. This denomination is due to the $g_{12} |v|^2 u $ and $g_{12} |u|^2 v $ terms, which depending on the value of $g_{12}$, favor solutions that coexist or that spatially separate. The segregation problem consists in studying the properties of the solutions when $g_{12}$ is positive and very large, since in this case, solutions tend to have disjoint supports. \\

The segregation of two-component BECs has been experimentally observed. The achievement of two-component BECs constitutes an important research subject in experimental physics, see for example \cite{HaJILA}, \cite{MaJILA98} or \cite{PaJILA}. Two-component BECs have been realized in several ways, using different types of particles: two different isotopes of the same atom, isotopes of two different atoms, or a single isotope in two different hyperfine states. The choice of the number of particles and the types of isotopes determine the value of $g_1$ and $g_2$. The value of $g_{12}$ can be altered by changing the hyperfine state of a portion of the particles, via applied magnetic or optical fields. This allows large and tunable values for $g_{12}$ (\cite{BlochDalZwer}, \cite{PaJILA}, \cite{ThalLENS}). Therefore, manipulating the values of the coupling strengths, it is possible to realize segregated two-component BECs with different spatial configurations (\cite{HaJILA}, \cite{PaJILA}). We refer to \cite{KaTsuUe} for the description of the physics.\\

In numerical analysis (\cite{KaTsuUe}, \cite{KaYaTsu}, \cite{MaAf}), several patterns have been observed for the supports of two-component BECs. In \cite{MaAf}, Mason and Aftalion perform a numerical analysis on the solutions of (\ref{eq:sc}). They classify all the possible configurations depending on $g_{12}$ and $g_1\neq g_2$. For the segregated case, they exhibit two main types of patterns: symmetry preserving and symmetry breaking ones. In the first case, one component is a disc centered in the minimum of $V$, and the other component is an annulus surrounding the disc. In the second case, both components are close to half balls.\\

In this paper, we address both segregation and symmetry breaking for two-component BECs in all $\R^2$. Our goal is to prove that when $g_{12}$ goes to infinity, the supports of $u_g$ and $v_g$ are disjoint; and that when $g_1$ and $g_2$ go to zero, they break the symmetry of the harmonic potential by approaching half-spaces. To our knowledge, the are no results in the mathematical literature about symmetry breaking for two-component BECs in the repulsive case.\\

Our method consists in two steps. First, we consider sequences $g^n=(g_1^n,g_2^n,g_{12}^n)$ such that $g_{12}^n$ goes to infinity and $g_1^n$ (respectively $g_2^n$) converges to some nonnegative limit $g_1$ (respectively $g_2$). We prove that the associated sequence of minimizers $(u_{g^n},v_{g^n})$ converges to a limiting pair, which minimizes the addition of the Gross-Pitaevskii energy of each component

\be \label{africahightech}
	\E_{g_1,g_2,\infty}(u, v) =  \frac12 \int_{\R^2}  \Big\{  |\nabla u|^2 + V |u|^2 + \frac12 g_1 |u|^4 \Big\} +  \frac12 \int_{\R^2}  \Big\{ |\nabla v|^2 + V |v|^2 + \frac12 g_2 |v|^4 \Big\} 
 \ee
 
 over
 
  \be \label{Blown}
	Y = \Big\{ (u,v) \in X \,;\, u \cdot v = 0 \Big\} \,,
 \ee
 
 the subset of $X$ of fully segregated pairs. Then, we study the ground states of $\E_{g_1,g_2,\infty}$ when $g_1$ and $g_2$ are both equal to zero, which results in an eigenvalue problem that leads to half-space geometry.\\
  
 Our first result is, \\

 \textbf{Theorem A.}  
\textit{ Let  $(u_{g^n} ,v_{g^n} )_{n \in \N}$ be a sequence of  minimizers of $ \E_{g^n} $ over $X$, with $g^n=(g_{1}^n,g_{2}^n,g_{12}^n) \in [0,c_0]^2 \times \R^+$ such that }

\be
	\lim_{n \to \infty}g_1^n=g_1 , \qquad \lim_{n \to \infty}g_2^n=g_2 \qquad \text{and} \qquad  \lim_{n \to \infty}g_{12}^n=\infty \,.
\ee

\textit{There exist a limiting pair $(u_\infty, v_\infty) \in Y$ such that, up to a subsequence,}
 
\begin{description}

\item[(i)] \textit{ $(u_{g^n} ,v_{g^n} )$ converges to $(u_\infty, v_\infty)$ weakly in $H^1_V(\mathbb{R}^2) \times H^1_V(\mathbb{R}^2)$ and in $C^0_{loc}(\mathbb{R}^2) \times C^0_{loc}(\mathbb{R}^2)$}.

\item[(ii)] \textit{$(u_\infty, v_\infty)$ minimizes $\E_{g_1,g_2,\infty}$ over $Y$ and satisfies weakly the system}

\be
 \left\{
	\begin{array}{rlll}
		-\Delta u + V u  + g_1 |u|^2 u  &=& \lambda u & \quad \text{ in }  \quad \{ |u_\infty| > 0\}  \\
		-\Delta v + V v  + g_2 |v|^2v  &=& \mu v & \quad \text{ in }  \quad \{ |v_\infty| > 0\} \\
		u \cdot v  &=& 0 & \quad \text{ in }  \qquad \, \,  \, \, \R^2 \,.
	\end{array}  \right. \label{eq:sc2}
\ee

\textit{Here $\lambda$ and $\mu $ are respectively the limits of the Lagrange multipliers $\lambda_n$ and $ \mu_n $, associated with $(u_{g^n},v_{g^n})$ by (\ref{eq:sc}).}

\end{description}

This is the analogue of the results of Wei and Weth in \cite{WeWe1} for solutions of the system (\ref{eq:sc}) in bounded domains and without trapping potential. We obtain the corresponding results by adapting the techniques in this papers to our setting. In \cite{WeWe1} (see also the work of Chang, Lin, Lin and Lin in \cite{clll}), the authors prove the segregation in the strongly coupled case and the local uniform convergence to a limiting pair solving the system (\ref{eq:sc2}).  In \cite{ctv2}, Conti, Terracini and Verzini study the equivalent of the energy $\E_{g_1,g_2,\infty}$ defined over bounded domains and without trapping potential. They prove the existence of minimizers and their Lipschitz regularity, and give extremality conditions in the form of a system of subsolution of elliptic equations. We emphasize that we address the problem that actually corresponds to the physical situation. In the previous mentioned works, no mathematical results are presented about the symmetry breaking for solutions of system (\ref{eq:sc}). \\

The main result of this article, and the improvement with respect to the previous works, is to prove the symmetry breaking and to give an accurate description of the behavior of the ground states, in the strongly coupled and weakly interaction case:\\

\textbf{Theorem B.}  
\textit{ Let  $(u_{g^n} ,v_{g^n} )$ be a sequence of  minimizers of $ \E_{g^n} $ over $X$, with $g^n=(g_{1}^n,g_{2}^n,g_{12}^n)$ such that $g_1^n \to 0$, $g_2^n \to 0$ and $g_{12}^n \to \infty$. Then, up to a subsequence, $u_{g^n}$ and $v_{g^n}$ converge in $C^0_{loc}(\R^2)$ respectively to $e^{i\theta_+}w_\nu^+$ and $ e^{i\theta_-}w_\nu^-$, where }

\be \label{heineken}
	w_\nu(x) = \frac2{\sqrt{\pi}} \, \omega \,  (x \cdot \nu) \,  e^{-\frac{\omega|x|^2}2} \,, \quad \nu \in S^1
\ee

\textit{is a second eigenfunction of the harmonic oscillator $-\Delta+\omega^2 |x|^2$ in $\R^2$, and $\theta_+$, $\theta_-$ are real constants.}\\

The limiting functions $w_\nu^+$ and $w_\nu^-$ are not radial since they are supported in the half-spaces $\{ x \cdot \nu >0\} $ and $\{ x \cdot \nu <0\}$. Theorem B says that in compacts domains of $\R^2$, the supports of $u_g$ and $v_g$ approach these half-spaces when $g_1 \to 0$, $g_2 \to 0$ and $g_{12} \to \infty$. The desired result about the symmetry breaking follows immediately:\\

\textbf{Corollary 1.1.} \textit{There are positive constants $g_0$ and $G$, such that if $(u_g,v_g)$ is a minimizer of $\mathcal{E}_{g}$ in $X$ with}

\ben
	\max \{ g_1,g_2 \} \leq  g_0 \qquad \text{and} \qquad g_{12} \geq G\,,
\een

\textit{ then $u_g$ and $v_g$ are not radially symmetric.}\\

Theorem B follows directly from the local uniform convergence stated in Theorem A, together with an accurate description of the segregated minimizers in the non interacting limit when $g_1=g_2=0$,  given by:\\

\textbf{Theorem C.} \textit{Let $(u_0,v_0)$ be a minimizer of $\E_{0,0,\infty}$ over $Y$. Then, \textbf{(i)} $u_0$ is supported in $H=\{x_1>0\}$ and $v_0$ in $H^c$ or rotated of them, and \textbf{(ii)} $u_0$ and $v_0$ are the positive and the negative parts of $w_\nu$, a second eigenfunction of the harmonic oscillator $-\Delta+\omega^2 |x|^2$ in $\R^2$ given by (\ref{heineken}).} \\    

We prove Theorem C using the results of Ehrhard in \cite{Ehr2} about the extremality properties of half-spaces. Theorem C says that in the limit case when $g_1=g_2=0$, the nodal set of a segregated two-component BEC is a straight line. We emphasize that in the case when $g_1$ and $g_2$ are small and positive, the exact configuration of both components is still unknown, but we expect the nodal set to be an infinite line with small curvature. We mention the work of Berestycki, Lin, Wei and Zhao in \cite{BeLinWeiZhao}, where they study the profile of the components near the interphase using blow-up techniques.\\  

The study of the geometric nature of the support of the segregated minimizers of $\E_{0,0,\infty}$ is an example of an optimal partition problem, or of a free boundary problem between several components. \\

The optimal partition problem consists in finding a partition of $\Omega \subset \R^n$ by $k$ disjoint open sets $\omega_1, \dots, \omega_k$, that minimizes a function of the groundstate energies in each component. This is an open problem in general, and only few results are known about the exact configurations of these optimal partitions. In \cite{CaffLin}, Caffarelli and Lin consider the problem of minimizing $\sum_{i=1}^k \lambda(\omega_i)$, where $\lambda(\omega)$ is the first Dirichlet eigenvalue of the Laplacian on $\omega$. For $\Omega$ bounded, they show the existence of classical solutions as well as the regularity of the interfaces. In \cite{HeHOTe}, Helffer, Hoffmann-Ostenhof and Terracini also consider the problem of minimizing $\max_{i=1,\dots,k} \mu(\omega_i)$, where $\mu(\omega)$ is the first Dirichlet eigenvalue of some Schr\"odinger operator on $\omega$. They study the relation between nodal domains, spectral partitions and spectral properties in a bounded domain. An important question is whether a minimal partition  is  nodal, that is, consisting in the nodal domains of an eigenfunction of the Dirichlet  realization of the Schr\"odinger operator in $\Omega$. This is the case for the second problem when $k=2$ and $\Omega$ is bounded (see \cite{ctv2} and \cite{HH}).\\

In Theorem C, we actually prove that the optimal partition, for the minimization of the addition of the energies of a system of two harmonic oscillators in $\R^2$, is composed of the supports of the positive and negative parts of its second eigenfunction.\\

The free boundary problem, which usually arises for competitive system when $g_{12}$ goes to infinity, has been studied by Caffarelli and Lin in \cite{CaffLin2}, for the singularly perturbed elliptic system $-\Delta u + g_{12}u^2v=0$ and   $-\Delta v+ g_{12}v^2u=0$. They prove the existence of a limiting pair, for which each component is a harmonic function in its domain of definition; and the $C^{0,\alpha}$ regularity of the nodal line $\{u=v=0\}$. They use Algrem's monotonicity formula, originally established in \cite{AltCaffFried}, which allows to prove H\"older uniform estimates for the solutions. See \cite{CaffKaLin}, \cite{ctv3} and \cite{NoTaTeVe}  for other works in this spirit. \\

A further related problem is the study of (\ref{eq:sc}) in the attractive case when $g_1$ and $g_2$ are both negative. This is certainly a very different problem, since in this case the associated energy may not be bounded from below (see for example \cite{LiWei}) and no ground states exist. Symmetry results have been proved in this case, when there is neither trapping potential nor mass constraints. In \cite{WeWe2}, Wei and Weth prove that in a system defined over all $\R^2$, they are infinitely many non radial solutions, and in \cite{WeWe4}, that in a system defined in the unit ball, for every $k \in \mathbb{N}^*$ there is a radially symmetric solution $(u,v)$ such that $u-v$ changes exactly $k$ times of sign in the radial variable. We also mention the work of Liu in \cite{Liu}, where he does not prescribe the constraint of the $L^2$-norm of each component being one, and therefore in the limit obtains that only one component remains. This is another form of segregation, since the model allows one component to disappear in the strong interacting case, when $g_1$ and $g_2$ go to infinity.\\

\textbf{Sketch of proofs}\\

We now explain the outline of the paper and the main ideas of the proofs. \\

In Section \ref{segregation}, we prove Theorem A and the segregation of the minimizing pairs $(u_g,v_g)$. We first remark that for all $g \in \R_+^3$ and $(u,v) \in Y$, $\mathcal{E}_g(u,v) =  \mathcal{E}_{g_1,g_2,\infty}(u,v)$, so $\E_g(u_g,v_g) \leq \mathcal{E}_{g_1,g_2,\infty}(u,v)$, and there are positive constants $c_1$, $c_2$ and $c_3$ such that for every $g_1\geq0$ and  $g_2\geq0$

\be
	\mathcal{E}_g(u_g,v_g) \leq c_1+c_2g_1+c_3g_2 \,,
\label{barking}
 \ee

for all $g_{12}\geq0$. \\

\textit{Sketch of the proof of Theorem A:} Using (\ref{barking}), the energy of $(u_{g^n},v_{g^n})$ (and also the Lagrange multipliers $\lambda_{g^n}$ and $\mu_{g^n}$) is uniformly bounded. Hence, there exists $(u_\infty, v_\infty) \in X$ which is the weak limit of $(u_{g^n},v_{g^n})$ in $H^1_V \times H^1_V$ and the strong limit in $L^2  \times L^2 $. We prove that $(u_\infty, v_\infty)$ minimizes $\E_{g_1,g_2,\infty}$ in $Y$ using energy estimates. In order to prove the $C^0_{loc}$ convergence, we follow the ideas Wei and Weth in \cite{WeWe1}, where they prove the equicontinuity of solutions of system (\ref{eq:sc}) in bounded domains and without trapping potential. We show that the proof of Theorem 1.1(a) therein works in our setting. The reason of this, is that the proof consists in a rescaling of the solutions, which yields in a limit problem over all $\R^2$. Rescaling $u_{g^n}$ and $v_{g^n}$ identically, we get the same limit problem, so the equicontinuity holds, which gives the local uniform convergence. To show that $(u_\infty, v_\infty)$ satisfies the system (\ref{eq:sc2}), we first prove some estimates for minimizing pairs $(u_g,v_g)$ of $\E_g$: there are $C_2>0$ and $C_3>0$ such that, $\|u_g\|_\infty , \|v_g\|_\infty \leq C_2$ and  $\|\nabla u_g\|_\infty , \| \nabla v_g\|_\infty \leq C_3 \sqrt{g_{12}}$, for any $g \in [0,c_0]^2 \times \R^+$ with $g_{12}$ large enough. Using these, we show in Proposition \ref{pr:1} that there is $C_4>0$, such that for every $\eps>0$ and every $\eta>1$, $|v_g| \leq C_4 \, g_{12}^{-\eta}$ in $\{ \inf_g u_g >\eps\} $ and  $|u_g| \leq C_4 \, g_{12}^{-\eta}$ in $\{ \inf_g v_g >\eps\} $, for $g_{12}$ large enough depending on $\eps$ and $\eta$. This implies that the functions $g^n_{12}|v_{g^n}|^2u_{g^n}$ and $g^n_{12}|u_{g^n}|^2v_{g^n}$ converge weakly to zero respectively in $ \{  u_\infty >0\}$ and $ \{  v_\infty >0\}$, and the system (\ref{eq:sc2}) is satisfied. \\

\textbf{Remark 1.2.} We expect  the derivatives of $u_\infty$ and $v_\infty$ to be not continuous trough the nodal line $\{ u_\infty =v_\infty=0\}$, and the system (\ref{eq:sc2}) to be solved only in the interior of the supports of $u_\infty$ and $v_\infty$. In \cite{KaYaTsu} and \cite{MaAf}, we find simulations supporting this idea. In addition, in the case when $g_1=g_2$, we conjecture that the difference of the two components is a smooth function. Indeed, as we explain in Remark 3.6, in this case, we expect $u_\infty-v_\infty$ to solve a homogeneous elliptic equation, which after standard elliptic regularity arguments, yields in smoothness for $u_\infty-v_\infty$. \\

In Section \ref{propsegr}, we study the properties of fully segregated BECs. We show that the minimizers of $\E_{g_1,g_2,\infty}$ in $Y$ are locally Lipschitz continuous, and that the nodal set has empty interior. The local Lipschitz continuity is an important result because later in the proof of Theorem C, we need the limiting function to be locally Lipschitz continuous. The result about the nodal line of the segregated minimizers is used in the proof of Theorem C. We stress here that the space $Y$ is not a manifold, so we cannot perform calculus of variations therein. We have to use other techniques to deal with minimizers of $ \mathcal{E}_{g_1,g_2,\infty}$ in $Y$, in order to get a system of equations allowing us to study their local properties. We will explain these techniques, based on the works of Conti, Terracini and Verzini in~\cite{ctv1} and~\cite{ctv2}. \\

\textbf{Remark 1.3.} All the results from Section \ref{segregation} and Section \ref{propsegr} do not depend on the specific form of the trapping potential, and they remain true when considering a potential V with polynomial growth at infinity, i.e., a real function $V$ such that

\ben
		  V(x) \ge c \, |x|^p  \qquad  \text{ in }  \mathbb{R}^2\setminus B_{r}(0) 
\een

for some $c>0$,  $r> 0$  and  $p> 0$.\\

In Section \ref{noninter}, we prove Theorem C and Theorem B. Theorem B follows directly from Theorem A and  Theorem C(ii). The key ingredient in the proof of Theorem C is the extremal property of the half-spaces $H_{a,\nu}=\{x \cdot \nu>a\}$ with respect to the Gaussian-Rayleigh quotient 

\ben
	F(u) =  \frac{\int   |\nabla u|^2 d\mu}{\int   | u|^2 d\mu}  \,.
\een

Here $a$ is a real number, $\nu \in S^1$ and $\mu$ is the Gaussian measure over $\R^2$. In \cite{Ehr2}, Ehrhard prove, roughly speaking, that if $\Lambda(S)$ is the infimum of $F$ over the functions vanishing outside of a Lebesgue measurable set $S$, and $H_{a,\nu}$ is a half-space with the same Gaussian measure as $S$, then 

\ben
	\Lambda(S) \geq \Lambda(H_{a,\nu}) \,,
\een

and the equality holds only when $S=H_{a,\nu}$ for some $\nu \in S^1$. \\

\textit{Sketch of the proof of Theorem C:} For the first assertion, we start noticing that the change of variables $ \tilde u(x) = \sqrt{2\pi} \, \alpha \, u(\alpha x) e^{\frac14 |x|^2}$, whit $\alpha=(2\omega)^{-\nicefrac12}$, gives

\ben
	\E_{0,0,\infty}(u,v) = \omega \, ( F(\tilde u) + F( \tilde v) +2) \,.
\een

Next, we first show in Lemma \ref{lemmEEF}  that if $(u_0,v_0)$ minimize $\E_{0,0,\infty}$ over $Y$, then 

\ben
	     F(\tilde u_0) = \Lambda(\tilde \U_0) \quad \text{and} \quad   F(\tilde u_0) = \Lambda(\tilde \V_0)
\een

where $\tilde \U_0$ and $ \tilde\V_0$ are respectively the supports of $\tilde u_0$ and $\tilde v_0$. Using Ehrhard's result, we establish then that

\ben
	     \tilde \U_0 =  H_{a,\nu} \quad \text{and} \quad    \tilde \V_0 =  H_{b,\nu'} 
\een

for some $a,b \in \R^2$ and $\nu, \nu' \in S^1$. Because $\tilde \U_0 $ and $\tilde \V_0  $ are disjoint, and because the nodal line has empty interior (see Proposition \ref{int_pts}), we get that $a=-b$ and $\nu'=\nu$.  Hence $\tilde \V_0=H_{a,\nu}^c$. Then, we use a result of Beckner, Kenig and Pipher (see \cite{CK}) saying that the mapping $a \mapsto \Lambda(H_{a,\nu})$ is convex, to prove that $a$ is equal to zero. The definition of the change of variables gives then that $supp \, u_0 = H_{0,\nu} $ and $supp \, v_0 = H^c_{0,\nu}$ for some $\nu \in S^1$. Finally, using standard arguments for one-component BECs, we prove that system (\ref{eq:sc2}) is uniquely solved in half-spaces. The second assertion of Theorem C comes then easily, remarking that the positive and the negative parts of a second eigenfunction of the harmonic oscillator solve (\ref{eq:sc2}) in half-spaces.

\section{The segregation limit}
\label{segregation}
		
	In this section we assume that $g_1$ and $g_2$ are bounded with respect to $g_{12}$, so there is $c_0>0$ such that 
 
 \be \label{sziget}
	\max\{ g_1, g_2\} \leq c_0 \,.
 \ee

Moreover, with out loss of generality, we assume that $u_g$ and $v_g$ are real positive functions over all $\R^2$. We are allowed to do this after the following result, which is standard for one component BECs (see \cite{Aflivre}):

 \begin{lemm}  \label{hercules} Each component of a minimizing pair of $\mathcal{E}_g$ over $X$ is, up to a complex multiplier of modulus one, a real positive smooth function.
  \end{lemm}
   \textbf{Proof.}  Every minimizing pair $(u_g,v_g)$ solves the system (\ref{eq:sc}), so using standard elliptic regularity and the strong maximum principle, $u_g$ and $v_g$ are non vanishing smooth complex functions. Thus, there are smooth real functions $\varphi_1$ and $\varphi_2$, such that $u_g=|u_g|e^{i\varphi_1}$ and $v_g=|v_g|e^{i\varphi_2}$. The diamagnetic inequality imply that $(|u_g|,|v_g|)$ is also a minimizer of $\E_g$ over $X$. We have then the equality $\E_g(|u_g|,|v_g|)=\E_g(u_g,v_g)$, which imply that  $\varphi_1$ and $\varphi_2$ are constants, and hence the result. \qed\\
 
We start showing uniform estimates on $\mathcal{E}_g $  and on $u_g$, $v_g$ and its derivatives.  We will use these to prove Theorem A and the segregation of $u_g$ and $v_g$ in Proposition \ref{pr:1}.  

 \subsection*{Estimates on minimizers}
 
 \begin{lemm} 
 \label{Emilie}
There are positive constants $C_0$,  $C_1$,  $C_2$ and $C_3$, such that  if $(u_g,v_g)$ is a minimizing pair of $\mathcal{E}_g$ over $X$, with $g=(g_1,g_2,g_{12}) \in [0,c_0] \times [0,c_0] \times \R_+$, and $\lambda_g$, $\mu_g$ are the associated Lagrange multipliers, then  \\
 
 \ba
 	\mathcal{E}_g(u_g,v_g) &\leq& C_0  \label{C0} \\
	0 < \lambda_g, \mu_g &\leq& C_1 \label{C1} \\
	\| u_g\|_\infty, \| v_g\|_\infty   &\leq& C_2 \label{C2} \\
	\| \nabla u_g\|_\infty, \| \nabla v_g\|_\infty   &\leq& C_3 \, \sqrt{g_{12}} \label{C3}
\ea

for every $g_{12}>1$.
 
 \end{lemm}
 
 \textbf{Proof.}  \textit{Proof of (\ref{C0}):} as we saw in (\ref{barking}), if $(u,v) \in Y$ then
 
 \ben
 	\E_g(u_g,v_g) \leq \mathcal{E}_g(u,v) =  \mathcal{E}_{g_1,g_2,\infty}(u,v) \,,
 \een

so there are positive constant $c_0$, $c_1$ and $c_2$, not depending on $g_1$, $g_2$ or $g_{12}$, such that  for every $g_1\geq0$ and  $g_2\geq0$

\ben
	\mathcal{E}_g(u_g,v_g) \leq c_0+c_1g_1+c_2g_2 = C_0 \,,
\een

for all $g_{12}\geq0$. We get then (\ref{C0}) after (\ref{sziget}).\\
 
 \textit{Proof of (\ref{C1}):}  Multiplying the first equation in (\ref{eq:sc}) by $u_g$ and then integrating over all $\mathbb{R}^2$ we get that
  
 \ben
	 \lambda_g  = \int  |\nabla u_g|^2 +V u_g^2 + g_1 u_g^4 + g_{12} u_g^2v_g^2 \,,
 \een
 
so after the mass constraint  $ \lambda_g >0$, and after (\ref{C0}), $ \lambda_g \leq  4 \, \mathcal{E}_{g}(u_g, v_g) \leq 4\, C_0=C_1$. The same argument is valid with $\mu_g $, which yields (\ref{C1}). \\

 \textit{Proof of (\ref{C2}):} Consider $x \in \R^2$ and $R>0$. Using (\ref{C0}), the mass constraint and the continuous embedding $H^1 \hookrightarrow  L^p$ for $p\in[2,\infty)$, for every ball $B=B_{2R}(x)$ there is a positive constant $C^{'}=C^{'}(p,R)$ such that

\be
\label{koeblen}
	  \|u_g\|_{L^p(B)} \leq C^{'} \|u_g\|_{H^1(B)} \leq C^{'} \sqrt{2 \, \E_g(u_g,v_g)+1} \leq C^{'} \sqrt{2 \, C_0+1} \,.
\ee

After (\ref{eq:sc}) we get

 \ben
	 -\Delta u_g \leq h_g \quad \text{in} \quad B \,
\een

with $h_g=\lambda_g u_g $. Using (\ref{C1}) together with (\ref{koeblen}), there is $C^{''}=C^{''}(q,R,g_1,g_2) >0 $ such that 

 \be
\label{koeblen2}
	  \|h_g\|_{L^{q/2}(B)} \leq C^{''} \,
\ee

 for every $q\in [4,\infty)$.\\
 
Using a local estimate for $H^1$ subsolutions of elliptic equations (see Theorem 8.17 in~\cite{GT}) there is $C^{'''}=C^{'''}(R,p,q)>0$ such that

\ben
	  \sup_{B_{R}(x)} u_g \leq C^{'''} \left( R^{-\nicefrac2p} \|u_g^+\|_{L^p(B)} + R^{2-\nicefrac4q} \|h_g\|_{L^{q/2}(B)} \right) \,.
\een
  
Fixing $R$, $p$ and $q$, we derive from (\ref{koeblen}) and (\ref{koeblen2}) that there is $C_2>0$ such that

\ben
	 u_g(x) \leq C_2
\een

for every $g_{12}>0$. The same argument is valid with $v_g$. Therefore, since $u_g$ and $v_g$ are positive, we get (\ref{C2}). \\  
  
  \textit{Proof of (\ref{C3}):}  we first  prove that $u_g$ and $v_g$ have polynomial decay at infinity. More precisely, we claim that for every $\alpha > 0$ there is $r_{\alpha}>0$ and $C_{\alpha}>0$ such that for all $g_{12}\geq0$,
  
\be 	\label{disconnectxxx}
	u_g(x)< \frac{C_{\alpha}}{|x|^{\alpha}} 
\ee

for all $x \in K_{\alpha}  =\mathbb{R}^2\setminus B_{r_{\alpha}}(0) $. For $\alpha$ fixed, take $r_{\alpha}^2 = \nicefrac{1}{2\omega^2} \big(C_1+\sqrt{C_1^2+4\alpha^2\omega^2}\big)$.  A straightforward calculation shows that 

\ben
	  f_{\alpha}(x) = C_2 \left( \frac{r_{\alpha}}{ |x|}\right)^{\alpha}
\een

is a supersolution of the first equation in (\ref{eq:sc}), 

\ben
	-\Delta f_{\alpha} + f_{\alpha} \,  ( V + g_1 f_{\alpha}^2+ g_{12} |v|^2  - \lambda_g )  \geq 0 \qquad \text{in} \qquad K_{\alpha}
\een

while

\ben
	  f_{\alpha}\Big|_{\partial K_{\alpha}} = C_2 \geq u_g \Big|_{\partial K_{\alpha}} .
\een

Now define $\psi = f_{\alpha}-u_g$  and suppose that $\psi$ is strictly negative somewhere in $K_{\alpha}$. Because $f_{\alpha}$ and $u_g$ are of class $C^2$ in $K_{\alpha} $ and go both to zero at infinity, $\psi$ must have a local minimum $x_0$ in $\mathring{K_{\alpha}}$ :  $u_g(x_0)>f_{\alpha}(x_0)$ and $D^2\psi(x_0)$ is positive defined, so $\Delta \psi(x_0) \geq 0$. Using this, and the fact that $f_{\alpha}$ is a super solution of the first equation in (\ref{eq:sc}) while $u_g$ is a solution, we have 

\ben
	   (V(x_0) + g_1 f_{\alpha}(x_0)^2 + g_{12}  v_g(x_0)^2 - \lambda_g )f_{\alpha}(x_0) \geq (V(x_0) + g_1 u_g(x_0)^2 + g_{12}  v_g(x_0)^2 - \lambda_g )u_g (x_0) \,.
\een

But our choice of $r_{\alpha}$ implies in particular that $V(x_0) - \lambda_g > 0$, so we get a contradiction with $\psi(x_0) <0$ and the claim is proved. Remark that the previous claim, together with (\ref{C2}) imply that $V u_g $ is uniformly bounded in $\mathbb{R}^2$ with respect to $g_{12}$. \\

To finish the proof, let $x \in \mathbb{R}^2$ and suppose that $g_{12}>1$. For $y \in B_{2}(0)$ define $\tilde{u}(y)=u(x+g_{12}^{-\nicefrac{1}{2}}y)$ and $\tilde{v}(x)=v(x_0+g_{12}^{-\nicefrac{1}{2}}y)$. We have

\ben
	  \Delta \tilde{u}(y) = g_{12}^{-1} \{ V(x+g_{12}^{-\nicefrac{1}{2}}y)\tilde{u}(y)+g_1\tilde{u}(y)^3-\lambda_g\tilde{u}(y) \} +  \tilde{u}(y) \tilde{v}^2(y) \,,
\een

so after (\ref{sziget}),  (\ref{C1}),  (\ref{C2}) and the claim there is a constant $c>0$ such that $| \Delta \tilde{u}(y) | \leq c$ for all $y \in B_{2}(0)$ and $g_{12}>1$. Using a H\"older estimate for the first derivative of $\tilde{u}$ (see Theorem 8.32 in~\cite{GT}) there is a constant $C>0$ such that
\ben
	 \|\nabla \tilde{u}\|_{L^{\infty}(B_1(0))} \leq  C(C_2+c) \,.
\een

We get then the result with $C_3=C(C_2+c)$ considering $\nabla \tilde{u} \, (0)$. \\  \qed \\

We now show the key ingredient in the Proof of Theorem A, the segregation of $u_g$ and $v_g$. This is a generalization of Proposition 2.1 in~\cite{clll} to positive solutions of (\ref{eq:sc}) defined in all $\mathbb{R}^2$. We have also used in the proof some ideas from~\cite{WeWe1}. \\

\begin{prop}
\label{pr:1}
Consider $(u_g,v_g)$ a sequence of minimizers of $\mathcal{E}_g$ over $X$ and define

\ben
	\U_\eps \equiv \{ x \in \R^2 \,;\, \inf_{g_{12}>0} u_g(x) \geq \eps \} \,, \quad \text{}	\quad \V_\eps \equiv \{ x \in \R^2 \,;\, \inf_{g_{12}>0} v_g(x) \geq \eps \}\,.
\een	
	
For any $\eps>0$ and $\eta>1$, there are $G_0>0$ and a positive constant $C_4$ such that 

\ben
	v_{g} \leq C_4 \, g_{12}^{-\eta} \quad \text{in} \quad \U_\eps \qquad \text{and} \qquad u_{g} \leq C_4 \,g_{12}^{-\eta}  \quad \text{in} \quad \V_\eps
\een

for every $g_{12}>G_0$.
\end{prop} 

\textbf{Proof.} Let $\eta>1$, $\eps>0$ and $x \in \U_\eps$. Define the numbers

\ben
	 \left. 
	\begin{array}{rclrcl}
		\eta_1 &=& \frac{8}{\epsilon} \, \eta & \hspace{1cm} \rho& \in &  (0,\frac{1}{2} \,   e^{-C_2^2})  \\
		s_{g} &=& \eta_1 \, g_{12}^{-\nicefrac{1}{2}} \, \ln{g_{12}} & \hspace{1cm} t_{g} &=& g_{12}^{-\rho}
	\end{array} \right.
\een

so $s_g < t_g$ for $g_{12}$ large enough depending on $\eta$ and $\eps$, and note $B=B_{s_g}(x)$. \\

Define also the function $h_g:(0,\infty) \rightarrow \mathbb{R}$ by

\ben
		h_g(r) = \frac{1}{2\pi r } \int_{\partial B_{r}(x)} u_{g}^2 \,ds \,,
\een

and notice that

\be
\label{rita}
		 2\pi r \, h'_g(r) = \int_{\partial B_{r}(x)} u_{g} \cdot \partial_{\nu} u_{g} \,ds \,.
\ee

After (\ref{C2}) we have that $0< h_g\leq C^2_2$. We claim the existence of $\xi_g \in (s_g,t_g)$ such that 

\be
		h'_{g}(\xi_g) \leq \frac{-1}{\xi_g \ln{\xi_g}} \,.
\label{eq:claim}
\ee

If not, we will get

\begin{eqnarray*}
		C_2^2 &>& h_g(t_g) -  h_g(s_g) > \int_{s_g}^{t_g}  \frac{-1}{r \ln{r}} \, dr  =  \ln{\left( \frac{\ln{s_g}}{\ln{t_g}} \right)} \\
		&=& \ln{\left( \frac{ \ln{(\eta_1 \,  \ln{g_{12}}) } - \frac{1}{2} \, \ln{g_{12}}} {- \rho \ln{g_{12}} }\right)} \xrightarrow{g_{12} \to \infty} \ln{(\frac{1}{2\rho})}
		\,,
\end{eqnarray*}

which contradicts the choice of $\rho$. \\

We have then

\ben
		\int_{ B} |\nabla u_{g}|^2 \,dx \leq \int_{B_{\xi_g}(x)} |\nabla u_{g}|^2 \,dx  = \int_{\partial B_{\xi_g}(x)} u_{g} \cdot \partial_{\nu} u_{g} \,ds - \int_{ B_{\xi_g}(x)}   \Delta u_{g} \cdot u_{g} \,dx \,.
\een

The first term of the right hand side can be estimated using (\ref{eq:claim}), and the second one using (\ref{eq:sc}), (\ref{C1}) and (\ref{C2}). We get

\ben
		\int_{ B} |\nabla u_{g}|^2 \,dx \leq -\frac{2\, \pi}{\ln{\xi_g}} +  C_1 \, C_2^2 \, \xi_g^2 \,,
\een

so there is a positive $C^{'}$ such that

\be
		\int_{ B} |\nabla u_{g}|^2 \,dx \leq \frac{C^{'}}{\ln{g_{12}}} 
\label{eq:nb}
\ee

for $g_{12}$ large enough. \\

Using Theorem 7.17 in~\cite{GT}, we have that for $p \in (2,3)$ and $\gamma=1-\frac{2}{p}$, there is $C_p>0$ such that

\begin{eqnarray*}
		\text{osc}_{B} \, u_{g}  &\leq& C_p \,s_g^{\gamma} \,  \| \nabla u_{g}\|_{\infty}^{(p-2)/p} \, \| \nabla u_{g}\|_2^{2/p} \,.
\end{eqnarray*}

Using (\ref{C3}), (\ref{eq:nb}) and the definition of $s_g$,  there is  $C^{''}>0$ such that

\begin{eqnarray*}
		\text{osc}_{B} \, u_{g} &\leq& C^{''} \, \eta_1^{\gamma} \, g_{12}^{-\frac{\gamma}{2} + \frac12 - \frac2p}\, (\ln{g_{12}})^{1-\frac{3}{p}} \\
		&=&  C^{''} \, \eta_1^{\gamma} \, (\ln{g_{12}})^{1-\frac{3}{p}} \,,
\end{eqnarray*}

so after (\ref{sziget}), $\text{osc}_{B} \, u_g \to 0 $ when $g_{12} \to \infty$. This implies that 

\ben
		u_{g} \geq \frac{\epsilon}{2} \qquad \text{in} \qquad B
\een

for $g_{12}$ large enough. \\

Using this last estimate, together with (\ref{eq:sc})  (\ref{C1}) and (\ref{C2}), we get 

\be
	 \left\{ 
	\begin{array}{ccl}
		-\Delta v_{g} \leq -g_{12} \frac{\epsilon^2}{16} \, v_{g} &  \text{in} & B
\\ \\
		v_{g} \geq 0   &  \text{in} & B
\\ \\
		v_{g} \leq C_2 &  \text{in} & \partial B
\\ \\

	\end{array} \right.
\label{eq:several}
\ee

for $g_{12}$ large enough. Hence, Lemma 4.4 in~\cite{ctv1} gives that exist a constant $C>0$ (not depending in $g$, $\eps$, $\eta$ or $x$),  such that  

\ben
		\|v_{g}\|_{\infty} \leq C \, C_2 \, e^{-\frac{s_g}2 \sqrt{g_{12} \frac{\epsilon^2}{16}}} \hspace{1cm} \text{in} \hspace{1cm} B_{\frac{s_g}{2}}(x) \,.
\een

Tacking $C_4=C\,C_2$, the definition of $s_g$ gives 

\ben
		v_g(x) \leq C_4  \, g_{12}^{-\eta} \,,
\een

for $g_{12}$ large enough depending on $\eta$ and $\eps$. The equivalent argument holds for $u_g$ in $\V_\eps$, which yields Proposition \ref{loclip}. \\ \qed \\

We have now all the tools to prove Theorem A.\\

\textbf{Proof of Theorem A.} Let $ (u_n,v_n)=(u_{g^n},v_{g^n})$ be a sequence of minimizing pairs of $\mathcal{E}_{g^n}$ in $X$ with $g_{1}^n \to g_1$, $g_{2}^n \to g_2$ and $g_{12}^n \to \infty$.\\

\textbf{(i)} After (\ref{C0})  the sequences $u_n$ and $v_n$ are bounded in $H^1_V$, so there exists $(u_\infty,  v_\infty) \in H^1_{V} \times H^1_{V} $ with (up to a subsequence)

\begin{eqnarray*}
	u_n &\rightharpoonup& u_\infty \hspace{.2cm} \text{in}\hspace{.2cm}  H^1_V   \\
	v_n &\rightharpoonup& v_\infty \hspace{.2cm} \text{in}\hspace{.2cm}  H^1_V \,,
\end{eqnarray*}

as $n \rightarrow \infty$. The compact embedding $H^1_V(\R^2) \hookrightarrow L^2(\mathbb{R}^2)$ gives the strong $L^2$ convergence, so $\|u_\infty\|_2=\|v_\infty\|_2=1$ and $(u_\infty,  v_\infty) \in X$.\\

On the one hand, we have that 

\be
\label{daniel}
	\|u_n v_n -u_\infty v_\infty\|_1 \leq   \|u_n\|_2  \|v_n -v_\infty\|_2 + \|v_\infty\|_2  \|u_n -u_\infty\|_2 = o(1) \,,
\ee	

so (up to a subsequence) $u_n v_n \to u_\infty v_\infty$ a.e. in $\mathbb{R}^2$. And on the other hand, after (\ref{C2}) and (\ref{disconnectxxx}), $u_n v_n \leq C_1 h \in L^2$. The Lebesgue dominated convergence theorem together with (\ref{C0}) gives then

\ben
	\|u_\infty v_\infty\|_2 = \lim_{n \to \infty} \|u_nv_n\|_2  \leq \lim_{n \to\infty} \frac {C_2}{g_{12}^n} =0 \,.
\een	

Hence, $u_\infty\cdot v_\infty=0$ a.e. in $\mathbb{R}^2$ and $(u_\infty, v_\infty) \in Y$.  \\

In order to prove the $C^0_{loc}$ convergence, we follow directly the ideas of Wei and Weth in \cite{WeWe1}. In Theorem 1.1(a) therein, they show that sequences of positive solutions of a class of competitive nonlinear elliptic systems in bounded domains are uniformly equicontinous. Their proof consists in a rescaling of the solutions and the domains, which yields in a limit problem over all $\R^2$. We will show that defining the rescaled functions of $u_{g^n}$ and $v_{g^n}$, we get the same limit problem, so the equicontinuity  holds. We recall that  $(u_n,v_n)$ satisfies the system (\ref{eq:sc}) and that after Lemma 2.2, the sequence is uniformly bounded in $H^1_V \times H^1_V$ and in $L^\infty \times L^\infty $.\\

Following the proof of Theorem 1.1(a) in  \cite{WeWe1}, if the sequence $(u_n,v_n)$ is not uniformly equicontinuous, there exists $\delta>0$ such that, without loss of generality, $u_n$ satisfies (up to a subsequence)

\ben	
	\inf \left\{ |x-y| \,;\, x,y \in \R^2 \,,\, |u_n(x)-u_n(y)| \geq 2\delta \right\}	 \to 0 \qquad \text{as } n \to \infty. 
\een

Then, since the $u_n$ functions are positives, (\ref{disconnectxxx}) implies that there are $x_n,y_n \in \R^2$ such that $r_n := |x_n-y_n| \to 0$ as $n \to \infty $, $d_n :=u_n(y_n) \geq \delta$ and $u_n(x_n) \geq d_n + \delta$.\\

Take $e_1=(1,0)$ and choose $A_n \in O(2)$ such that $A_ne_1=r_n^{-1}(y_n-x_n)$. We define the rescaled function $v_{i,n}: \R^2 \to \R^+$ by 

\ben
	v_{1,n}(x) = u_n(x_n+r_nA_ny)  \qquad \text{ and } \qquad v_{2,n}(x) = v_n(x_n+r_nA_ny) \,.
\een

Then, $v_{1,n}$ solves in $\R^2$

\be \label{medusa}
	 \left\{
	\begin{array}{rcl}
		-\Delta v_{1,n} &=& l_{1,n}v_{1,n} - r_n^2 \, g_{12}^n \, v_{1,n} \\
		 v_{1,n} &>& 0   \\
		 v_{1,n}(e_1) &=& d_n \,\, \geq  \,\, \delta \\
		 v_{1,n}(0) &\geq& v_{1,n}(e_1)+\delta \,.
	\end{array}  \right. 
\ee

Here $l_{1,n} (x)=  r_n^2 \left( V(x_n+r_nA_nx)+g_1^nv_{1,n}^2(x)-\lambda_n \right)$, which after (\ref{disconnectxxx}), (\ref{C1}) and (\ref{C2}) satisfies 

\be  \label{medusa2}
	l_{1,n} v_{1,n}  \to 0 \quad \text{in } L^\infty \qquad \text{ as } n \to \infty\,.
\ee

Moreover, after (\ref{C0}) $v_{1,n} $ is uniformly bounded in $H^1$, and Lemma \ref{pr:1} also applied for the sequence $(v_{1,n},v_{2,n})$. This last two properties, together (\ref{medusa}), (\ref{medusa2}), implies that $(v_{1,n},v_{2,n})$ satisfies  the same hypotheses as in the proof of  Theorem 1.1(a) in  \cite{WeWe1}. Hence, we obtain the same limit problem when $n$ goes to infinity, which following exactly the proof, yields a contradiction. The desired result then holds.\\

\textbf{(ii)} For the first assertion, let $(\tilde{u},\tilde{v})$ be any pair in $Y$. Then

\be \label{twilight1}
	\mathcal{E}_{g_1^n,g_2^n,\infty}(\tilde{u},\tilde{v})=\mathcal{E}_{g^n}(\tilde{u},\tilde{v}) \geq \mathcal{E}_{g^n}(u_n,v_n) \geq \mathcal{E}_{g_1^n,g_2^n,\infty}(u_n,v_n) \,.
 \ee
 
Since the pair $(u_n,v_n)$ satisfies the uniform bounds (\ref{disconnectxxx}) and (\ref{C2}), the $L^2$ convergence implies the $L^4$ convergence. This, together with (\ref{twilight1}) and the weak lower semicontinuity of the ${H^1_V}$ norm, gives 

\ban
     \mathcal{E}_{g_1,g_2,\infty}(\tilde{u},\tilde{v}) &=& \liminf_{n \to \infty} \mathcal{E}_{g_1^n,g_2^n,\infty}(\tilde{u},\tilde{v}) \\
	&\geq&  \liminf_{n \rightarrow \infty} \mathcal{E}_{g_1^n,g_2^n,\infty}(u_n,v_n) \\
	&\geq&  \mathcal{E}_{g_1,g_2,\infty}(u_\infty,v_\infty) \,,
 \ean

which imply the result since after (i), $(u_\infty,v_\infty) \in Y $. \\

For the second assertion, let $\varphi$ be a $C^{\infty}$ function supported in $K \subset\subset \{ u_\infty>0\}$. Multiplying  the first equation on (\ref{eq:sc}) by $\varphi$ and then integrating, we get

\be
\label{feels}
	\int_K \nabla \varphi \cdot \nabla u_n + \varphi \, ( V u_n  + g_1^n  u_n^3 -  \lambda_n   u_n) = - \int_K  \varphi \,  g_{12}^n \, v_n^2 \, u_n \,.
\ee	

Using the weak convergence of $u_n$ to $u_\infty$, the left hand side of (\ref{feels}) tends  to 

\ben
	\int \nabla \varphi \cdot \nabla u_\infty + \varphi \, (V u_\infty  + g_1  u_\infty^3 -  \lambda   u_\infty )
\een	
 
 with $\lambda$ the limit (up to a subsequence) of $\lambda_n$, which exists because of (\ref{C1}). \\

After (i), $u_n$ converges uniformly to $u_\infty$ in $K$. Hence, $K \subset \subset \{ u_\infty>2\eps\}$ for some $\eps > 0$, and there exists $N >0$ such that 

\ben
	K \subset \{\inf_{n>N} u_{g^n} \geq \eps \} \,.
\een

Thus, Proposition \ref{pr:1}, together with (\ref{C2}), implies that $ g_{12}^n \,v_n^2 \, u_n$ converge uniformly to zero in $K$, so the right hand side of (\ref{feels}) tends to zero as $n \to \infty$. Hence, 
 
\ben
	\int \nabla \varphi \cdot \nabla u_\infty + \varphi \, (V u_\infty  + g_1  u_\infty^3) =  \lambda \int \varphi u_\infty 
\,.
\een

The same argument is valid with $v_\infty$, which yields the result.\\

\qed

\section{Properties of fully segregated two-component BECs}
\label{propsegr}
		In this section we prove some properties of fully segregated two-component BECs. The results in here will be used to prove the symmetry breaking in the limit case when $g_1=g_2=0$. We start with the local Lipschitz continuity of minimizers of $\E_{g_1,g_2, \infty}$ in $Y$.

\subsection*{Local Lipschitz continuity}

\begin{prop}
 \label{loclip}
If $(u,v)$ is a nonnegative real minimizer $\E_{g_1,g_2,\infty}$ over $Y$, then $u$ and $v$ are  locally Lipschitz continuous in $\mathbb{R}^2$.
\end{prop}

To prove this proposition, we first see in Lemma \ref{subsol} that each component of a minimizing pair of $\E_{g_1,g_2, \infty}$ is a weak subsolution of a Gross-Pitaevskii equation, and that the difference of both components is a weak solution of a non homogeneous elliptic equation. We derive local uniform estimates with respect to $g_1$ and $g_2$ for the minimizing pairs. These estimates imply that the minimizers satisfy some extremality conditions. We conclude using the following result of Conti, Terracini and Verzini (\cite{ctv2}, Theorem 8):

\begin{theo}\textbf{(Conti-Terracini-Verzini)} Let $\Omega$ be a bounded regular set of $\mathbb{R}^2$, $M\geq0$ and $ w_1,w_2 \in H^1(\Omega)$ such that $w_1 \geq 0$,  $w_2 \geq 0$ and $w_1 \cdot w_2 =0$. If $w_1$ and $w_2$ satisfy

\begin{equation*}
  		-\Delta w_1 \leq M \,, \hspace{1cm}  -\Delta w_2 \leq M  \hspace{.5cm}  \text{and} \hspace{.5cm}   -M \leq -\Delta (w_1-w_2)  \leq M \,, 
\end{equation*}

then they are both Lipschitz continuous in the interior of $\Omega$.
\label{CTV}
\end{theo}

 \begin{lemm}
 \label{subsol}
Let $g_1\geq$ and $g_2\geq0$. If $(u,v)$ is a nonnegative minimizer of $\E_{g_1,g_2,\infty}$ over $Y$ then

\begin{equation}
\label{eq:ine1}
  		-\Delta u  + V u + g_1 u^3 \leq \lambda  u  \,, \hspace{1cm}    \text{} \hspace{.5cm}   -\Delta v + V v  + g_2 v^3  \leq \mu v \hspace{.5cm} \hspace{.05cm} 
\end{equation}

and 

\begin{equation}
\label{eq:e1}
  		-\Delta (u-v)  + V (u-v) + g_1u^3 - g_2 v^3=  \lambda u- \mu v 	
\end{equation}

in $\mathcal{D}'(\mathbb{R}^2)$, where  $\lambda = e_1(u)$ and $\mu = e_2(v)$ with

 \begin{eqnarray*}
	 e_i(w) &=&  \int    |\nabla w|^2 + V | w|^2 + g_i |w|^4  
 \end{eqnarray*}

for $i=1,2$.
 \end{lemm}

\textbf{Proof.} \textit{Proof of (\ref{eq:ine1}):} Arguing by contradiction, suppose that  

\begin{equation*}
		 \int \nabla u  \cdot \nabla \phi +  ( V u   + g_1 u^3  -  \lambda u ) \, \phi > 0
 \end{equation*}
 
for some $0\leq \phi \in C^{\infty}_0(\mathbb{R}^2)$. \\

For $t \in (0,1)$, define a new test function as :
 
 \begin{eqnarray*}
	 (w_1,w_2) &=& \left( \frac{(u-t\phi)^+}{\|(u-t\phi)^+\|_2} \,,\, v \right).
 \end{eqnarray*}

Where $u^+=\max(u,0)$ and $u^-=\max(-u,0)$. In the rest of the proof the $o(\cdot)$ notation will mean with respect to the $t \to 0$ limit.\\

Since $\{ (u-t\phi)^+  >0 \} \subset \{ u>0 \}$, $(u-t\phi)^+ \cdot v =0$ a.e. in $\mathbb{R}^2$, and $(w_1,w_2) \in Y$. \\

Using $f^2= (f^+)^2 +(f^-)^2$, $\|u\|_2=1$ and $0 \leq (u-t\phi)^-\leq t\phi$, we compute

\begin{eqnarray*}
		\|(u-t\phi)^+\|_2^2 &=& 1 +  \int \left( [(u-t\phi)^+]^2 - u^2\right) \\
					&=&  1 -  \int  \left( 2t\phi u + [(u-t\phi)^-]^2 - t^2 \phi^2 \right) \\
					&=& 1 -  \int  2t\phi u  + o(t)  \,,
 \end{eqnarray*}

so
 
 \begin{equation*}
		\frac1{\|(u-t\phi)^+\|_2^{2}} = 1 +  \int  2t\phi u  + o(t)   \quad \text{and} \quad \frac1{\|(u-t\phi)^+\|_2^{4}} = 1 +  \int  4t\phi u  + o(t) \,.
 \end{equation*}

\vspace{.5cm}

The difference between the energies is then

\begin{eqnarray}
	\E_{g_1,g_2,\infty}(w_1,w_2)-\E_{g_1,g_2,\infty}(u,v) =   \frac12  \int\left( |\nabla (u-t\phi)^+|^2-|\nabla u|^2 \right)  &+& \int  t\phi u  \cdot \int |\nabla (u-t\phi)^+|^2  \nonumber \\
	+ \frac12\int V \left( [(u-t\phi)^+]^2-u^2\right)  &+&  \int  t\phi u  \cdot \int V (x)|(u-t\phi)^+|^2  \label{eq:diffenergy}\\
	+ \frac14\int g_1 \left( [(u-t\phi)^+]^4-u^4\right)  &+& \int  t\phi u  \cdot \int g_1  |(u-t\phi)^+|^4 + o(t)\,.\nonumber 
\end{eqnarray}

We note that 

\begin{eqnarray*}
	\int \frac12 \left(  |\nabla (u-t\phi)^+|^2 -|\nabla u|^2 \right)  & \leq&   \int  \frac12 \left( |\nabla (u-t\phi)|^2 -|\nabla u|^2 \right)  \\
	&=& -t \int \nabla \phi \cdot \nabla u + o(t) \,,
\end{eqnarray*}

so (\ref{eq:diffenergy}) becomes

\begin{eqnarray*}
	\E_{g_1,g_2,\infty}(w_1,w_2)-\E_{g_1,g_2,\infty}(u,v)  &\leq& -t \int  \nabla u \cdot \nabla \phi  +  V u \phi + g_1 u^3 \phi - e_1((u-t\phi)^+)  \, u \phi    + o(t) \,.
\end{eqnarray*}

Using Lebesgue dominated convergence theorem, we have that $e_1((u-t\phi)^+) \to e_1(u)$ when $t \to 0$, so  

\begin{eqnarray*}
	\E_{g_1,g_2,\infty}(w_1,w_2)-\E_{g_1,g_2,\infty}(u,v)  &\leq& -t \int  \nabla u \cdot \nabla \phi  + ( V u  + g_1 u^3  - \lambda u ) \, \phi   + o(t) \,.
\end{eqnarray*}

Hence, for $t$ small enough we get the contradiction

\begin{equation*}
	\E_{g_1,g_2,\infty}(w_1,w_2) - \E_{g_1,g_2,\infty}(u,v) <0 \,.
\end{equation*}

Using the same arguments with $v$ and $\mu$, the inequalities in (\ref{eq:ine1}) are proved. \\

\textit{Proof of (\ref{eq:e1}):} Define $\hat{u} = u-v$ and suppose that  

\begin{equation*}
		  \int \nabla \hat{u} \cdot \nabla \phi + (  V \hat{u}+ g_1 u^3 -  g_2 v^3 -  \lambda u  +\nu  v )\phi  < 0
 \end{equation*}
 
for some $0\leq \phi \in C^{\infty}_0(\mathbb{R}^2)$. \\

For $t \in (0,1)$, define a new test function as :
 
 \begin{equation*}
	 (w_1,w_2) = \left( \frac{(\hat{u} +t\phi)^+}{\|(\hat{u} +t\phi)^+\|_2} \,,\, \frac{(\hat{u} +t\phi)^-}{\|(\hat{u} +t\phi)^-\|_2} \right).
 \end{equation*}
 
 As before, we compute 

 \begin{equation*}
		\frac1{\|(\hat{u}+t\phi)^+\|_2^{2}} = 1 -  \int  2t\phi u  + o(t) \,,  \qquad \qquad   \frac1{\|(u+t\phi)^+\|_2^{4}} = 1 -  \int  4t\phi u + o(t)  \,,
 \end{equation*}

 \begin{equation*}
		\frac1{\|(\hat{u}+t\phi)^-\|_2^{2}} = 1 +  \int  2t\phi v  + o(t)  \quad \text{ and } \quad \frac1{\|(u+t\phi)^-\|_2^{4}} = 1 +  \int  4t\phi v  + o(t)  \,.
 \end{equation*}
 
\vspace{.5cm}
 
Using that $u\cdot v=0$,  the difference between the energies is

\begin{eqnarray*}
	 \E_{g_1,g_2,\infty}(w_1,w_2)-\E_{g_1,g_2,\infty}(u,v) =  \frac12  \int \left( |\nabla (\hat{u}+t\phi)|^2-|\nabla \hat{u}|^2 \right) &+& V \left( | (\hat{u}+t\phi)|^2 - |\hat{u} |^2 \right)  \\
	 \hspace{2cm} + \, \frac14 \int g_1 \left( | (\hat{u}+t\phi)^+|^4 - |u|^4\right) &+& g_2   \left( | (\hat{u}+t\phi)^-|^4 - |v|^4\right)  \\
	- t \int \big( \, u \,  e_1((\hat{u}+t\phi)^+) &-&  v  \, e_2((\hat{u}+t\phi)^-)  \, \big) \,   \phi  + o(t)   \,.
\end{eqnarray*}

 Hence,
 
 \begin{eqnarray*}
	 \E_{g_1,g_2,\infty}(w_1,w_2)-\E_{g_1,g_2,\infty}(u,v)  &=&  t \ \int \nabla  \hat{u}  \cdot  \nabla \phi  + V \hat{u}  \phi  +  (g_1u^3 - g_2 v^3) \phi \\
	&-& t \int  \big( \,u \,  e_1((\hat{u}+t\phi)^+) -  v  \, e_2((\hat{u}+t\phi)^-) \, \big)  \, \phi + o(t)     \,.
\end{eqnarray*}

Using the same argument as before, we see that  $e_1((\hat{u}+t\phi)^+) - e_1(u)=o(1)$ and  $e_2((\hat{u}+t\phi)^-) - e_2(v)=o(1)$, so

 \begin{eqnarray*}
	 \E_{g_1,g_2,\infty}(w_1,w_2)-\E_{g_1,g_2,\infty}(u,v)  &=&  t \ \int \nabla  \hat{u}  \cdot  \nabla \phi  +( V \hat{u}   + g_1u^3 - g_2 v^3) \,, \phi  \\
	&-& t \int   \big( u \,  \lambda-  v  \, \mu\big) \,  \phi + o(t)     \,.
\end{eqnarray*}

And again, for $t$ small enough we get

\begin{equation*}
	\E_{g_1,g_2,\infty}(w_1,w_2) - \E_{g_1,g_2,\infty}(u,v) <0 \,,
\end{equation*}

a contradiction.\\

We have proved the inequality

\begin{equation*}
		 \int \nabla (u-v) \cdot \nabla \phi +  V (u-v) \phi  + (g_1 u^3 -  g_2 v^3) \phi-  (\lambda u  - \nu_{\infty} v )\phi \geq 0   \,.
 \end{equation*}

Using the same arguments with $\hat{v}=v-u$ we get 

\begin{equation*}
		 \int \nabla (v-u) \cdot \nabla \phi +  V (v-u) \phi  + ( g_2 v^3-g_1 u^3 ) \phi - ( -\lambda u  + \nu_{\infty} v )\phi \geq 0
 \end{equation*}

for every  $0\leq \phi \in C^{\infty}_0(\mathbb{R}^2)$. Equality (\ref{eq:e1}) is then proved. \\
 \qed\\

\textbf{Proof of Proposition \ref{loclip}:}  Let $\Omega$ be any bounded set of $\R^2$. After (\ref{eq:ine1}), $u$ and $v$  are respectively $H^1$ subsolutions of $-\Delta u=\lambda u$ and $-\Delta v=\mu v$. Arguing as in the proof of (\ref{C2}) $u$ and $v$ are uniformly bounded in $\Omega$ with respect to $g_1$ and $g_2$, and since they minimize $\E_{g_1,g_2,\infty}$, $\lambda$ and $\mu$ are also uniformly bounded. There are then positive $M_1, M_2=\O(g_1,g_2)$ such that 

\begin{equation*}
	-\Delta u \leq M_1 \quad \text{and} \quad -\Delta  v \leq M_2 \qquad \text{in} \quad \Omega \,,
\end{equation*}

so after (\ref{eq:e1}) and the previous estimates, there are positive $M_3, M_4=\O(g_1,g_2)$ such that

\begin{equation*}
	 -M_3 \leq -\Delta (u-v) \leq M_4  \qquad \text{in} \quad \Omega \,.
\end{equation*}

Theorem \ref{CTV} with $M=\max \{M_1,M_2,M_3,M_4\}$ implies then that $u$ and $v$ are Lipschitz continuous in $\Omega$ and the result is proved.  \\ 
\qed

\subsection*{The nodal set}

We now prove that the nodal set has empty interior. We do this by following an idea of Chang et all. in \cite{clll}. The point is that if the nodal set has a ball $B$ contained in it, then $u$ can be extended to a solution $\tilde u$ of an elliptic equation in $\text{supp} \, u  \cup B$, but then, after the strong maximum principle and the mass constraint, $\tilde u $ cannot vanish in the interior of its support. \\

\begin{prop}
 \label{int_pts}
If $(u,v)$ is a nonnegative real minimizer of $\E_{g_1,g_2,\infty}$ over $Y$, then the nodal set $\{ x \in \mathbb{R}^2 \,;\, u(x)=v(x)=0\}$ has no interior points.
\end{prop}

\textbf{Proof.} After Proposition \ref{loclip}, we know that $u$ and $v$ are locally Lipschitz functions, so $\U = \{ x \in \R^2 \,;\, u(x)>0 \}$ and $\V = \{ x \in \mathbb{R}^2 \,;\, v(x)>0 \}$ are open regular sets. Define $\tilde \U= \mathbb{R}^2 \setminus \V$ and suppose that the nodal line has an interior point. Then $\U \subsetneq \tilde{\U}$ and the Lebesgue measure of $ \U$ is less than the Lebesgue measure of  $\tilde \U$. Let $\tilde w$ be the minimizer of

\begin{equation*}
	 \E_{g_1,0,0}(w) = \int	_{\tilde{\mathcal{U}}}  \left\{ \frac{1}{2}  |\nabla w|^2 +\frac{1}{2}V | w|^2 + \frac{1}{4} g_1 |w|^4 \right\}
\end{equation*}

over the functions $ w$ in $H^1_0( \tilde \U )$ such that $\int_{ \tilde \U }  w^2 =1$. \\

Is clear that $(\tilde w,v) \in Y$, so after $\E_{g_1,g_2,\infty} (\tilde w,v) \geq \E_{g_1,g_2,\infty} (u,v) $ we get $\E_{g_1,0,0}(u) \leq \E_{g_1,0,0}(\tilde w)$. This imply that $u$ solves $ -\Delta u + (V + g_1|u |^2-\lambda ) u =0$, with $\lambda$ defined as in Lemma \ref{subsol}. Therefore, by elliptic regularity, $u$ is a $C^2$ function in $\tilde \U$, and using the strong maximum principle, $u\equiv0$ because it vanish in the interior of $\tilde \U$. This contradicts the mass constraint, so the nodal line has no interior points. \\ \qed \\

\textbf{Remark 3.6.} As we said in the introduction, we expect the derivatives of $u$ and $v$ to be not continuous trough the nodal line. In the case when $g_1=g_2$, we also expect the difference of the two components to be a smooth function. Indeed, after Proposition \ref{subsol}, if $(u,v)$ is a nonnegative real minimizer of $\E_{g_1,g_2,\infty}$ over $Y$, then $u-v$ solves the elliptic equation (\ref{eq:e1}) in $\R^2$. When $g_1=g_2$, we expect  the Lagrange multipliers $\lambda$ and $\mu$ to be equal, and equation (\ref{eq:e1}) to be homogeneous. Standard elliptic regularity theory implies then that $u-v$ is a $C^\infty$ function over all $\R^2$. These properties are verified when $g_1$ and $g_2$ are both equal to zero: in Theorem C(ii) we show that in this case, $u$ and $v$ are respectively the positive and negative parts of  a second eigenfunction of the harmonic oscillator $L$ in $\R^2$, so $u - v \in C^{\infty}(\R^2)$,  $\lambda=\mu$, and there is a jump of the derivatives of $u$ and $v$ trough the nodal line.\\

\section{The non interacting limit}
\label{noninter}

	In this section we study the minimizers $(u_0,v_0)$ of $\E_{0,0,\infty}$ over $Y$. We prove Theorem C, i.e., that the two components are supported in complementary half-spaces meeting at zero, and that they are respectively the positive and negative parts of a second eigenfunction of the harmonic oscillator $-\Delta+\omega^2|x|^2$ in $\R^2$. \\

For $a \in \R$ and $\nu\in S^1$ we define the half space

\ben
	H_{a,\nu} = \{ x \in \R^2 \,;\, x \cdot \nu> a \}	
\een

and we write $H_{a} = H_{a, (0,1)}$. The main idea in the proof of Theorem C is the extremal property of half-spaces with respect to the Gaussian-Rayleigh quotient

\ben	
	F(f) = \frac{\int |\nabla f|^2 d\mu }{\int | f|^2 d\mu } \,.
\een 

Here $\mu$ is the Gaussian measure in $\R^2$, which density with respect to the Lebesgue measure is given by

\ben
	d\mu(x)=  \frac1{2\pi} \, e^{-\frac12 |x|^2 }  dx \,.
\een

We remark that the invariance of the Gaussian measure with respect to rotations gives that for every $\nu,\nu'  \in S^1$

\be \label{cahier}
	\L(H_{a, \nu'}) = \L(H_{a,\nu}) = \L(H^c_{-a, \nu}) \,,
\ee

and that since $ \bar\R \ni a \mapsto \mu(H_{a, \nu}) $ is an increasing function, for every Lebesgue measurable set $S$ there is a real $a$ such that $\mu(S) =  \mu(H_{a, \nu}) $.\\

For a non empty open  $S$, we define $\F(S)$ as the class of nonnegative non zero functions absolutely continuous on lines with support in $S$, and $\L$ by

\ben	
	\L(S) = \inf_{f \in \F(S)} F(f) \,.
\een

In \cite{Ehr2} and \cite{Ehr}, Ehrhard study isoperimetric inequalities in Gauss spaces and introduce the Gaussian symmetrization, a variant of classical symmetrizations used to solve isoperimetric problems, such as the principal frequency of a membrane or the torsional rigidity of a bar (see for example \cite{BroZie}, \cite{PoSze} or \cite{Sarv}).  In \cite{Ehr}, Ehrhard prove that among all subsets  with prescribed Gaussian measure, half-spaces have minimal $\L$:

\begin{theo} \label{Ehrhard} \textbf{(Ehrhard)}
Let $S$ be a non empty open subset of $\R^2$ and $a \in \R$ such that $\mu(S) = \mu(H_{a}) $. Then, 
	 
\ben
	 \L(S)  \geq  \L(H_{a}) 
\een
	
and the equality holds if and only if $S=H_{a,\nu}$ for some $\nu \in S^1$. Moreover, the infimum in $\L(H_{a})$ is attained by some $f \in \F(H_{a})$. 

\end{theo}

We recall that the trapping potential is given by (\ref{jones}), so after (\ref{africahightech}) the energy is 

\ben
	\E_{0,0,\infty}(u, v) =  \frac12 \int_{\R^2}  \Big\{  |\nabla u|^2 + \omega^2 |x|^2 |u|^2  \Big\} +  \frac12 \int_{\R^2}  \Big\{ |\nabla v|^2 +  \omega^2 |x|^2 |v|^2  \Big\} \,,
 \een
 
and that the class of minimization $Y$ is given by (\ref{Blown}).\\

To prove Theorem C, we first perform a change of variable in order to deal with the minimization problem in a different setting. For $(u,v) \in Y$ define 

\be \label{chgvar}
	\u(x) = \sqrt{2\pi} \, \alpha \, u(\alpha x) \, e^{\frac14|x|^2} \quad \text{ and } \quad \v(x) = \sqrt{2\pi} \, \alpha \, v(\alpha x) \, e^{\frac14|x|^2} 
\ee

with $\alpha = (2\omega)^{-1/2}$, and 

\ben
	\Y = \{ (\u, \v) \,;\, (u,v) \in Y\} \,. 
\een

A direct computation gives 

\be \label{eff}
	\E_{0,0,\infty}(u, v) = \omega \, \big( F(\u) + F(\v) +2 \big) \,.
\ee

\begin{lemm} \label{lemmEEF}
A nonnegative pair $(u,v)$ minimizes $\E_{0,0,\infty}$ over $Y$ if and only if  $(\u,\v)$ minimize $F(\u)+F(\v)$ over $\Y$. Moreover, in this case $\u$ minimizes $F$ over $\F(\tilde \U)$ and $\v$ minimizes $F$ over $\F(\tilde \V)$, where $\tilde \U=supp \, \u$ and $\tilde \V = supp \,  \v$.
\end{lemm}

\textbf{Proof.} The first assertion is immediate after (\ref{eff}). If $(u,v)$ is a nonnegative minimizer of $\E_{0,0,\infty}$ over $Y$, then after Proposition \ref{loclip} $u$ is in $C^{0,1}_{loc}(\R^2)$, so it is absolutely continuous on lines, and because of the mass constraint u is not identically zero. After (\ref{chgvar}) the same properties hold for $\u$, so $\u \in \F(\tilde \U)$. For every $\w \in \F(\tilde \U)$ with finite Gaussian-Rayleigh quotient, $(\nicefrac w{\|w\|_2},v) \in Y$. Thus, (\ref{eff}) gives $F(\u) \leq F(\w)$. The same argument is valid with $\v$, so the second assertion is proved.\\

We are now able to prove Theorem C. For the first part, we use the ideas of Beckner, Kenig and Pipher in Section 2.4 of \cite{CK}.\\   

\textbf{Proof of Theorem C(i).} To lighten the notation we write $u=u_0$, $\U=supp \, u$, $\tilde \U=supp \, \u$ and the analogous for $v_0$. The diamagnetic inequality imply that $(|u|,|v|)$ is also a minimizer of $\E_{0,0,\infty}$ over $Y$, so we  suppose, without loss of generality, that $u$ and $v$ are nonnegative real functions. \\

\textit{Step 1: $\tilde \U=H_{a,\nu}$ and  $\tilde \V = H_{b,\nu'}$.} Suppose that  $\tilde \U$ or $\tilde \V$ is not a half-space. Then, after Lemma \ref{lemmEEF} and Theorem \ref{Ehrhard} there are real numbers $a,b$ such that $\mu(\tilde \U) = \mu(H_{a}) $, $\mu(\tilde \V) = \mu(H_{b}) $ and 

\be \label{f2l4}
	F(\u) + F(\v) = \L(\tilde \U) +  \L(\tilde \V) >  \L(H_a) +  \L(H_b) \,.
\ee

We claim that $a+b\geq0$. First, since $\mu(\tilde \U \cap \tilde \V)=0$ we have 

\be \label{muamub}
	\mu(H_a) + \mu(H_b) \leq 1 \,,
\ee

which imply that $a$ and $b$ cannot be both negative. We suppose then without loss of generality that 

\be \label{twoinq}
	b<0\leq a \qquad \text{ and } \qquad a+b<0 \,.
\ee

The fact that  $\mu(H_a) + \mu(H_{-a}^c) =1$, together with (\ref{muamub}), implies that $\mu(H_b) \leq \mu(H_{-a})$, which contradicts (\ref{twoinq}). The claim is then proved. \\  

The inequality $a+b\geq0$ imply that $H_a \cap H_{-b}^c = \emptyset$. Hence, for every pair $(\w_1,\w_2) \in \F(H_a) \times \F(H_{-b}^c) $, $(\nicefrac{\w_1}{ \|w_1\|_2},\nicefrac{\w_2}{ \|w_2\|_2}) \in \Y$. After (\ref{eff}) and (\ref{f2l4}) we obtain 

\ben 
	F(\w_1) + F(\w_2) \geq F(\u) + F(\v) >  \L(H_a) +  \L(H_b) \,.	
\een

Minimizing $F$ in the previous inequality with respect to $\w_1 \in \F(H_a)$ and with respect to $\w_2 \in \F(H_{-b}^c) $,  and considering (\ref{cahier}), we obtain the contradiction $\L(H_a) +  \L(H_b)  >  \L(H_a) +  \L(H_b)$, so Step 1 is proved.\\

\textit{Step 2: $a=-b$ and  $\nu=\nu'$.} Since $\u \cdot \v = 0$, $\mu(H_{a,\nu} \cap H_{-b,\nu'})=0$, which imply that the boundaries of $H_{a,\nu} $ and $H_{-b,\nu'} $ must be parallel, i.e., $\nu=\nu'$ and $a \geq -b$. Moreover, if $a>b$, then the nodal set $\{ \u=\v=0\}$ has a non empty interior, which after (\ref{chgvar}) contradicts Proposition \ref{int_pts}. We have shown that

\ben
	\tilde \U=H_{a,\nu} \qquad \text{ and } \qquad \tilde \V = H^c_{a,\nu} 
\een

for some $a \in \R$, $\nu \in S^1$.\\ 

\textit{Step 3: $a=0$.} First, the monotonicity of the first eigenvalue of the Dirichlet problem with respect to the domain gives that $a \mapsto \L(H_{a})$ is an increasing function. Moreover, the same argument of Theorem 2.4.5 in \cite{CK} (see also Theorem 6.2 in \cite{BL}) gives that it is a convex function. Considering (\ref{cahier}) we derive 

\be \label{rabbatre2}
	 \frac{\L(H_a) + \L(H^c_a)}2 =  \frac{\L(H_a) +  \L(H_{-a} )}2 \geq \L (H_0)  = \frac{\L(H_0) + \Lambda(H_0^c)}2 \,.
\ee   

Suppose now that $a\neq0$ and consider $(\w_1,\w_2) \in \F(H_0) \times \F(H_0^c) $. The same argument used in \textit{Step 1}, together with (\ref{rabbatre2}), gives

\ben
	F(\w_1) + F(\w_2) > \L(H_a) +  \L(H_a^c) \geq  \L(H_0) +  \L(H_0^c) \,.	
\een

After Theorem \ref{Ehrhard}, the infima in $ \L(H_0) $ and $ \L(H^c_0) $ are attained, so again, minimizing $F$ in the previous inequality with respect to $\w_1 \in \F(H_0)$ and with respect to $\w_2 \in \F(H_{0}^c) $, we obtain the contradiction $ \L(H_0) +  \L(H_0^c) >  \L(H_0) +  \L(H_0^c)$. We have proved that 

\ben
	\tilde \U=H_{0,\nu} \qquad \text{ and } \qquad \tilde \V = H^c_{0,\nu} \,,
\een

which considering (\ref{chgvar}), gives

\ben
	 \U=H_{0,\nu} \qquad \text{ and } \qquad  \V = H^c_{0,\nu} 
\een

for some $\nu \in S^1$.\\   \qed   \\

We now give the proof of the second part of Theorem C, this is, that every minimizing pair of $\E_{0,0,\infty}$ over $Y$ is of the form $\sqrt{2} \, ( e^{i \theta_+} \, w_\nu^+, e^{i \theta_-} \, w_\nu^-)$ for some $\nu \in S^1$ and $\theta_+, \theta_- \in \R$.\\

The second eigenvalue of  the harmonic oscillator $-\Delta+\omega^2 |x|^2$ over $\R^2$ has multiplicity 2. An orthogonal base, with respect to the $L^2$ product, of the associated spectral space is given by

\ben
	 \eta_1(x)  =    c_\omega \, x_1   \,    e^{-\frac{\omega |x|^2}2}  \qquad  \text{ and } \qquad     \eta_2 (x) =   c_\omega \,  x_2   \,    e^{-\frac{\omega |x|^2}2} \,
\een

with $c_\omega= \sqrt{ \nicefrac2{\pi}} \, \omega$. Every function in the spectral space, with $L^2$ norm equal to 1, is the of the form \\

\be \label{naturarekin}
	w_\nu(x) = c_\omega \,  (  x \cdot \nu ) \,    e^{-\frac{\omega |x|^2}2}
\ee

for some $\nu \in S^1$. \\

\textbf{Proof of Theorem C(ii).} We write $u_0=u$, $v_0=v$ and $H_{0,\nu}=H_\nu$. After Theorem A(ii) and Theorem C(i), there is $\nu \in S^1$ such that $u$ solves    

 \be \label{est}
	\hspace{1cm} \left\{ 
	\begin{array}{rcccr}
		-\Delta u+ \omega^2 |x|^2 u &=& \lambda  \, u& \text{in} & H_\nu
\\ 
		u &=& 0    &   \text{in} & \partial  H_\nu
\\
		\int_{H_\nu} u^2&=& 1    &  & 	\end{array} \right. 
\ee

with $\lambda = \int_{H_\nu} |\nabla u |^2 + \omega^2 |x|^2 u^2 $, and $v$ solves the equivalent problem in $H^c_\nu$. \\

The same argument as in Lemma \ref{hercules} imply that $u = e^{i\theta_+} |u| $ in $H_\nu$ and $v = e^{i\theta_-} |v| $ in $H_\nu$, with $|u|$ and $|v|$ positive and $\theta_+,\theta_- \in \R$. We claim now that there is uniqueness for the modulus of $u$ in problem (\ref{est}). Suppose that there are two positive solutions $u_1$ and $u_2$ of (\ref{est}) respectively with $\lambda_1$ and $\lambda_2$. Suppose that $\lambda_1 \leq \lambda_2$ and  define $h=u_1/u_2$ in $H_\nu$. We have that

 \begin{equation}
	 \nabla ( u_2^2 \, \nabla h) = (\lambda_2-\lambda_1) \, u_2^2 \, h \,,
\label{rihanna}
\end{equation}

and using the mass constraint that 

 \begin{equation*}
	 \int_{H_\nu}  u_2^2 h(h-1)  = \frac12 \int _{H_\nu} u_2^2 (h-1)^2 \,.
\end{equation*}

Multiplying (\ref{rihanna}) by $h-1$, then integrating over $H_\nu$ and performing an integration by parts we find

\begin{equation*}
	 \int_{H_\nu}   u_2^2  |\nabla h |^2 +  \frac12 (\lambda_2-\lambda_1) \, u_2^2 (h-1)^2 = 0 \,,
\end{equation*}
  
 which imply that $h\equiv cte$. The mass constraints imply then that $h\equiv 1$, so the claim is proved. The equivalent result is valid for v in $H_\nu^c$.\\

Finally, a direct computation shows that 
 
\ben
	u_\nu (x) = \sqrt 2 \, c_\omega \,  (  x \cdot \nu )^+    \,    e^{-\frac{\omega |x|^2}2} \qquad \text{ and } \qquad v_\nu (x) =  \sqrt 2 \, c_\omega    \,   ( x \cdot \nu )^-    \,    e^{-\frac{\omega |x|^2}2}
\een
are positive solutions of problem (\ref{est}) respectively in $H_{\nu}$ and $H^c_{\nu}$, so $u=e^{i\theta_+}u_\nu$ and $v=e^{i\theta_-}v_\nu$ for some $\nu \in S^1$ and $\theta_+, \theta_- \in \R$.  \\ \qed

We finish this article by given the proof of Theorem B. \\

\textbf{Proof of Theorem B.} Let $(u_{g^n} ,v_{g^n} )$ be a sequence of  minimizing pairs of $ \E_{g^n} $, with $g_1^n \to 0$, $g_2^n \to 0$ and $g_{12}^n \to \infty$. After Theorem A, $u_{g^n}$ (respectively $v_{g^n}$) converges locally uniformly to $u_0$ (respectively $v_0$), where $(u_0, v_0)$ minimizes $\E_{0,0,\infty}$ over $Y$. Theorem C(ii) gives then that $u_0=e^{i\theta_+}w_\nu^+$ and $v_0= e^{i\theta_-}w_\nu^-$ for some $\nu \in S^1$ and $\theta_+, \theta_- \in \R$. \\ \qed

\vspace{.5cm}	
	
The author would like to thank C. Kenig, J. Wei, A. Aftalion and S. Terracini for their help and suggestions, and P. Mason for sharing with me his physical insights.

\printindex

\bibliographystyle{acm}
\bibliography{bib}

\begin{thebibliography}{10}

\bibitem{Aflivre}
{\sc Aftalion, A.}
\newblock {\em Vortices in Bose-Einstein Condensates}, vol.~67 of {\em Progress
  in Nonlinear Differential Equations and Their Applications}.
\newblock Birkh\"auser, 2006.

\bibitem{AltCaffFried}
{\sc Alt, H.~W., Caffarelli, L.~A., and Friedman, A.}
\newblock Variational problems with two phases and their free boundaries.
\newblock {\em Trans. Amer. Math. Soc. 282}, 2 (1984), 431--461.

\bibitem{BeLinWeiZhao}
{\sc Berestycki, H., Lin, T.-C., Wei, J., and Zhao, C.}
\newblock On phase-separation model: Asymptotics and qualitative properties.
\newblock {\em Preprint\/} (2010).

\bibitem{BlochDalZwer}
{\sc Bloch, I., Dalibard, J., and Zwerger, W.}
\newblock Many-body physics with ultracold gases.
\newblock {\em Rev. Mod. Phys. 80}, 885 (2008), 885--964.

\bibitem{BL}
{\sc Brascamp, H.~J., and Lieb, E.~H.}
\newblock On extensions of the {B}runn-{M}inkowski and {P}r\'ekopa-{L}eindler
  theorems, including inequalities for log concave functions, and with an
  application to the diffusion equation.
\newblock {\em J. Funct. Anal. 22\/} (1976), 366--389.

\bibitem{BroZie}
{\sc Brothers, J., and Ziemer, W.}
\newblock Minimal rearrangements of {S}obolev functions.
\newblock {\em J. Reine Angew. Math. 384\/} (1988), 153--179.

\bibitem{CK}
{\sc Caffarelli, A., and Kenig, C.~E.}
\newblock Gradient estimates for variable coefficient parabolic equations and
  singular perturbation problems.
\newblock {\em Amer. J. Math. 120}, 2 (1998), 391--439.

\bibitem{CaffKaLin}
{\sc Caffarelli, L.~A., Karakhanyan, A.~L., and Lin, F.-H.}
\newblock The geometry of solutions to a segregation problem for nondivergence
  systems.
\newblock {\em J. Fixed Point Theory Appl. 5\/} (2009), 319--351.

\bibitem{CaffLin}
{\sc Caffarelli, L.~A., and Lin, F.-H.}
\newblock An optimal partition problem for eigenvalues.
\newblock {\em J. Sci. Comput. 31}, 1-2 (2007), 5--18.

\bibitem{CaffLin2}
{\sc Caffarelli, L.~A., and Lin, F.-H.}
\newblock Singularly perturbed elliptic systems and multi-valued harmonic
  functions with free boundaries.
\newblock {\em J. Amer. Math. Soc. 21}, 3 (2008), 847--862.

\bibitem{clll}
{\sc Chang, S.-M., Lin, C.-S., Lin, T.-C., and Lin, W.-W.}
\newblock Segregated nodal domains of two-dimensional multispecies
  {B}ose-{E}instein condensates.
\newblock {\em Physica D: Nonlinear Phenomena 196}, 3-4 (2004), 341--361.

\bibitem{ctv1}
{\sc Conti, M., Terracini, S., and Verzini, G.}
\newblock Asymptotic estimates for the spatial segregation of competitive
  systems.
\newblock {\em Adv. Math. 195}, 2 (2005), 524--560.

\bibitem{ctv3}
{\sc Conti, M., Terracini, S., and Verzini, G.}
\newblock On a class of optimal partition problem related to the
  {F}u$\check{\text{c}}$\'ik spectrum and to the monotonicity formulae.
\newblock {\em Calc. Var. Partial Differential Equations 22}, 1 (2005), 45--72.

\bibitem{ctv2}
{\sc Conti, M., Terracini, S., and Verzini, G.}
\newblock A variational problem for the spatial segregation of
  reaction-diffusion systems.
\newblock {\em Indiana Univ. Math. J. 54}, 3 (2005), 779--815.

\bibitem{Ehr2}
{\sc Ehrhard, A.}
\newblock Sym\'etrisation dans l'espace de {G}auss.
\newblock {\em Math. Scand. 53}, 2 (1983), 281--301.

\bibitem{Ehr}
{\sc Ehrhard, A.}
\newblock In\'egalit\'es isop\'erimetriques et int\'egrales de {D}irichlet
  gaussiennes.
\newblock {\em Ann. Sci. \'Ecole Norm. Sup. (4) 17}, 2 (1984), 317--332.

\bibitem{GT}
{\sc Gilbarg, D., and Trudinger, N.~S.}
\newblock {\em Elliptic Partial Differential Equations of Second Order}.
\newblock Springer, 1977.

\bibitem{HaJILA}
{\sc Hall, D.~S., Matthews, M.~R., Ensher, J.~R., Wieman, C.~E., and Cornell,
  E.~A.}
\newblock Dynamics of component separation in a binary mixture of
  {B}ose-{E}instein condensates.
\newblock {\em Phys. Rev. Lett. 81}, 8 (1998), 1539--1542.

\bibitem{HH}
{\sc Helffer, B., and Hoffmann-Ostenhof, T.}
\newblock Converse spectral problems for nodal domains.
\newblock {\em Mosc. Math. J. 7}, 1 (2007), 67--84.

\bibitem{HeHOTe}
{\sc Helffer, B., Hoffmann-Ostenhof, T., and Terracini, S.}
\newblock Nodal domains and spectral minimal partitions.
\newblock {\em Ann. Inst. H. Poincar{\'e} Anal. Non Lin{\'e}aire 26}, 1 (2009),
  101--138.

\bibitem{KaTsuUe}
{\sc Kasamatsu, K., Tsubota, M., and Ueda, M.}
\newblock Vortices in multicomponent {B}ose-{E}instein condensates.
\newblock {\em Int. J. Mod. Phys. B 19}, 1835 (2005).

\bibitem{KaYaTsu}
{\sc Kasamatsu, K., Yasui, Y., and Tsubota, M.}
\newblock Macroscopic quantum tunneling of two-component {B}ose-{E}instein
  condensates.
\newblock {\em Phys. Rev. A 64}, 053605 (2001).

\bibitem{LiWei}
{\sc Lin, T.-C., and Wei, J.}
\newblock Ground state of {N} coupled nonlinear {S}chr{\"o}dinger equations in
  $\mathbb{R}^n$, $n\leq3$.
\newblock {\em Comm. Math. Phys. 255}, 3 (2005), 629--653.

\bibitem{Liu}
{\sc Liu, Z.}
\newblock Phase separation of two-component {B}ose-{E}instein condensates.
\newblock {\em J. Math. Phys. 50}, 10 (2009), 102104.

\bibitem{MaAf}
{\sc Mason, P., and Aftalion, A.}
\newblock Classification of the ground states and topological defects in a
  rotating two-component {B}ose-{E}instein condensate.
\newblock {\em Phys. Rev. A 84}, 3 (2011), 033611.

\bibitem{MaJILA98}
{\sc Matthews, M.~R., Hall, D.~S., Jin, D.~S., Ensher, J.~R., Wieman, C.~E.,
  and Cornell, E.~A.}
\newblock Dynamical response of a {B}ose-{E}instein condensate to a
  discontinuous change in internal state.
\newblock {\em Phys. Rev. Lett. 81}, 2 (1998), 243--247.

\bibitem{NoTaTeVe}
{\sc Noris, B., Tavares, H., Terracini, S., and Verzini, G.}
\newblock Uniform {H}{\"o}lder bounds for nonlinear {S}chr\"odinger systems
  with strong competition.
\newblock {\em Comm. Pure Appl. Math. 63}, 3 (2010), 267--302.

\bibitem{PaJILA}
{\sc Papp, S.~B., Pino, J.~M., and Wieman, C.~E.}
\newblock Tunable miscibility in a dual-species {B}ose-{E}instein condensate.
\newblock {\em Phys. Rev. Lett. 101}, 4 (2008), 040402.

\bibitem{PoSze}
{\sc P\'olya, G., and Szeg\H{o}, G.}
\newblock {\em Isoperimetric inequalities in mathematical physics}.
\newblock Annals of Mathematics Studies, no. 27. Princeton University Press,
  1951.

\bibitem{Sarv}
{\sc Sarvas, J.}
\newblock Symmetrization of condensers in $n$-space.
\newblock {\em Ann. Acad. Sci. Fenn. Ser. A I}, 522 (1972).

\bibitem{ThalLENS}
{\sc Thalhammer, G., Barontini, G., Sarlo, L.~D., Catani, J., Minardi, F., and
  Inguscio, M.}
\newblock Double species {B}ose-{E}instein condensate with tunable interspecies
  interactions.
\newblock {\em Phys. Rev. Lett. 100}, 21 (2008), 210402.

\bibitem{WeWe2}
{\sc Wei, J., and Weth, T.}
\newblock Nonradial symmetric bound states for a system of coupled
  {S}chr\"odinger equations.
\newblock {\em Atti Accad. Naz. Lincei Cl. Sci. Fis. Mat. Natur. Rend. Lincei
  (9) Mat. Appl. 18}, 3 (2007), 279--293.

\bibitem{WeWe1}
{\sc Wei, J., and Weth, T.}
\newblock Asymptotic behaviour of solutions of planar elliptic systems with
  strong competition.
\newblock {\em Nonlinearity 21}, 2 (2008), 305--317.

\bibitem{WeWe4}
{\sc Wei, J., and Weth, T.}
\newblock Radial solutions and phase separation in a system of two coupled
  schr\"odinger equations.
\newblock {\em Arch. Ration. Mech. Anal. 190}, 1 (2008), 83--106.

\end{thebibliography}

\end{document}